 \documentclass{jnst} 
\usepackage[tablesfirst,notablist]{endfloat} 
\usepackage{graphicx}

\title{Phenomenological level density model with hybrid parameterization of deformed and spherical state densities}
\author{Naoya Furutachi\thanks{Corresponding author. Email: naoya.furutachi@riken.jp} \thanks{Present address: {RIKEN Nishina center, 2-1 Hirosawa, Wako, Saitama 351-0198, Japan}}, Futoshi Minato and Osamu Iwamoto}
\inst{Nuclar Data Center, Japan Atomic Energy Agency, Tokai-mura, Naka-gun, Ibaraki 319-1195, Japan}
\abst{
A phenomenological level density model that has different level density parameter sets 
for the state densities of the deformed and the spherical states, 
and the optimization of the parameters using experimental data 
of the average s-wave neutron resonance spacing are presented.  
The transition to the spherical state from the deformed one is described using 
the parameters derived from a microscopic nuclear structure calculation.
The nuclear reaction calculation has been performed by the statistical model using the present level density.
Resulting cross sections for various reactions with the spherical, 
deformed and transitional target nuclei show a fair agreement with the experimental data, 
which indicates the effectiveness of the present model.  
The role of the rotational collective enhancement in the calculations of those cross sections is also discussed.

}
\kword{nuclear level density; nuclear data; statistical model; neutron spectrum}

\begin{document}
\maketitle
\newpage
\section{Introduction}
The level density (LD) is a key ingredient in the nuclear reaction calculation using the statistical model.
The accuracy of the calculated nuclear reaction observables 
for various reaction channels relies on the LD, and therefore a number of theoretical works 
employing phenomenological \cite{2,3,4,25} or microscopic models \cite{5,6,7,8,9,10,11} 
have been devoted to achieve a reliable LD. 
While the microscopic models are basically free from adjustable parameters and suitable to predict 
LDs of nuclei away from the stability line,  
the phenomenological models that have analytical formula and adjustable parameters 
are still useful to calculate LDs of nuclei around the stability line for the practical applications. 
Generally, the reliability of the phenomenological models is ensured with the experimental information of
excitation energies and spin-parity of the low-lying discrete states, and the average of the s-wave neutron resonance spacing $D_0$. 

  
One of the key effects for LD is the enhancement due to the collective nuclear excitations. 
It is theoretically indicated that the collective rotational excitation 
brings an extremely large enhancement on the LD, which amounts to 10$\sim$100 magnitude 
at the neutron threshold energy of stable nuclei \cite{1,25}. 
In spite of its huge effect, the phenomenological LD models without the explicit treatment of the collective 
enhancement have been successfully applied to the nuclear reaction calculations for practical uses, for example, 
the LD model of Gilbert and Cameron \cite{2} without the collective enhancement \cite{3}
has been mainly used in the statistical model calculation of the neutron induced reaction under 20 MeV 
for the nuclear data evaluation of Japanese Evaluated Nuclear Data Library (JENDL) \cite{32}.
The reason why such a LD model does not cause serious problems in nuclear reaction calculations 
is conjectured that the collective enhancement is effectively taken into account in LD parameters, 
if they are optimized using the experimental $D_0$ \cite{3,4}. 

Actually, such an effective LD model works well for the optimization of the asymptotic level density parameter to reproduce $D_0$. 
Koning et al. \cite{4} have derived the global LD parameter systematics
for the several LD models with and without explicit treatment of the collective enhancement. 
As for the Fermi-Gas based models, both the collective and the effective LDs have a similar precision for the reproduction of $D_0$ to each other. 

It is noted that, besides the phenomenological models discussed here, 
the importance of the explicit treatment of the collective excitation is rather obvious in the microscopic LD calculations 
using Hartree-Fock plus Bardeen-Cooper-Schrieffer (HF+BCS) theory with the partition function method \cite{6,7}, 
and Hartree-Fock-Bogiliubov (HFB) theory with the combinatorial method \cite{8,9,10,11}.
All these studies treated the collective excitation explicitly, and found a fair agreement between the calculated $D_0$ and the experiments. 
These results indicate that if the intrinsic state densities are calculated without the parametrization, the collective enhancements are naturally required.

The role of the explicit treatment of the collective enhancement in phenomenological LD models 
can be discussed from nuclear reaction calculations.  
Koning et al. \cite{4} have applied the the effective and the collective LD models to  
systematic calculations of the nuclear reactions.
The calculated cross sections are systematically different between them for various reaction channels.
The difference is expected to be more significant in a nuclear reaction at a higher incident energy, 
because the asymptotic behaviors of LD models with and without the collective enhancement are quite different.
Actually, the important role of the collective enhancement in the cross section calculation 
for the projectile fragmentation with a relativistic incident energy have been reported \cite{19}. 



However, there remains problems in the description of the collective enhancement in phenomenological models.
One is the fading of the collective enhancement as a function of the excitation energy.
Although there are some theoretical investigations about the fading of the collective enhancement \cite{20}, 
it is difficult to confirm their validity directly from the experiments, because it is expected that
there is a finite mean deformation even with the excitation energy of several tens of MeV \cite{20} for well-deformed nuclei.
In addition to that, it is also difficult to describe the rotational collective enhancement for nuclei in the transitional region, 
because the interaction between the single-particle states and the collective states plays a significant role in this case.

Our aim in this paper is to present a reliable LD necessary for the precise calculation of 
nuclear reaction observables using the statistical model.
For this purpose, we propose a new phenomenological model based on the LD model of Gilbert and Cameron \cite{2}, 
in which the state densities of the deformed and spherical states have different level density parameters. 
The optimization of the parameters are performed by fitting the experimental $D_0$ 
with distinction between deformed and spherical nuclei.   
The LDs of the deformed and the spherical states are smoothly connected 
by the damping function, in analogy with the way used in the microscopic calculations based on HF+BCS and HFB \cite{6,7,8,9,10,11}. 
The fading of the rotational collective enhancement is effectively described in this way. 
Since there is no direct experimental information about the fading of the rotational collectivity, 
we utilized the microscopic nuclear structure calculation to determine the parameters in the damping function.
By the composition of the deformed and spherical states, 
the transitional state may be also effectively taken into account.

In this study, much attention is paid on the effectiveness of the present LD model for the actual nuclear reaction calculations.
We use CCONE code \cite{14} to calculate the cross sections, which are compared with the experimental data.
At the same time, we investigate the role of the explicit treatment of the collective enhancement in nuclear reactions.  
 
This paper is organized as follows. In Sec. \ref{theory}, the formulation of the present LD model,
the optimization procedure of the level density parameters, 
and the microscopic nuclear structure calculation are presented. 
In Sec. \ref{res}, first the characteristics of the present LD is discussed, 
then the results of the nuclear reaction calculations are shown.  
Sec. \ref{sum} summarizes this work.   

\section{Formulation} \label{theory}

We present a new phenomenological LD model that is described with the LDs 
of the deformed and the spherical states connected by the damping function 
in a similar way to that used in the microscopic calculations based on HF+BCS and HFB \cite{6,7,8,9,10,11}.  
By optimizing the level density parameters for the deformed and the spherical states separately,
reliable LDs for both deformed and spherical nuclei are expected to be achieved.
We call the present model as the hybrid model to distinguish from the existing phenomenological collective models.

\subsection{hybrid level density model}
The LD of the present hybrid model $\rho_{\rm h}$ is described with the LD of the spherical state $\rho_{\rm sph}$, 
and that of the deformed state $\rho_{\rm def}$,  
\begin{eqnarray} \label{eq_cond}
  \rho_{\rm h} (U,J) = \left\{ \begin{array}{ll}
    (1-f_{\rm dam}(E_x))\rho_{{\rm sph}}(U-E_{\rm def},J) + f_{\rm dam}(E_x)\rho_{{\rm def}}(U,J) &  (E_{\rm def} \geq E_{\rm cut}) \\
     & \\
    \rho_{{\rm sph}}(U,J) &  (E_{\rm def}<E_{\rm cut}),
  \end{array} \right.
\end{eqnarray}
which are smoothly connected by the damping function $f_{\rm dam}$,
\begin{eqnarray} \label{eq_fdam}
f_{\rm dam}(E_x) &=&  \frac{1}{1+e^{(E_x-E_{{\rm ts}})/d_{\rm e}}}, \hspace{2mm} d_e=C E_{\rm ts}. 
\end{eqnarray}
Here $E_x$, $U$ and $J$ are the excitation energy, the pairing corrected excitation energy 
and the total angular momentum of the nucleus.
In this formulation, the fading of the rotational collectivity is phenomenologically expressed by the 
transition from $\rho_{\rm def}$ to $\rho_{\rm sph}$. Since experimental information about the 
fading of the rotational collectivity is limited, we derived the parameters $E_{\rm def}$ and $E_{\rm ts}$  
that control this transition from the microscopic nuclear structure calculation, which is explained in Sec. \ref{mic}. 
The parameter $E_{\rm def}$ is defined as the energy difference between 
the deformed ground state and the minimum energy of the spherical state. 
If $E_{\rm def}$ is smaller than $E_{\rm cut}$, the level density is 
approximated with $\rho_{\rm sph}$. The parameter  $E_{\rm cut}$ is arbitrary fixed at 0.3 MeV. 
The parameter $E_{\rm ts}$ is the central energy of the transition, 
which is estimated utilizing information of the deformation at a finite temperature.
The width parameter $d_e$ of the damping function is phenomenologically determined 
supposing a linear dependence on $E_{\rm ts}$ with the adjustable parameter $C$.
The detailed discussion for the parameter $C$ is given in Sec. \ref{opt}.
The pairing corrected effective excitation energy $U$ is, 
\begin{eqnarray} \label{pair}
U &=& E_x-2\Delta \hspace{1cm} \textrm{for even-even nuclei} \nonumber \\
&=& E_x-\Delta \hspace{1cm} \textrm{for odd nuclei} \nonumber \\
&=& E_x \nonumber \hspace{1cm} \textrm{for odd-odd nuclei} \\
\Delta &=& 11/\sqrt{A}.
\end{eqnarray}

The functions $\rho_{\rm sph}$ and $\rho_{\rm def}$ are 
described by the phenomenological Fermi-gas model with the level density parameters $a_{\rm s}$ and $a_{\rm d}$, respectively, 
\begin{eqnarray}
\rho_{\rm sph}(U,J) = R_{\rm s}(U,J) \frac{\omega_{\rm s}(U)}{\sqrt{2\pi}\sigma_{\rm s}}, \nonumber \\
\rho_{\rm def}(U,J) = K_{\rm rot}R_{\rm d}(U,J) \frac{\omega_{\rm d}(U)}{\sqrt{2\pi}\sigma_{\rm d}}, \\
\omega_{\rm s,d}(U)=\frac{\sqrt{\pi}}{12}\frac{\exp(2\sqrt{a_{\rm s,d}U})}{a_{\rm s,d}^{1/4}U^{5/4}}, 
\end{eqnarray}
here $R_{\rm s,d}(U,J)$ are the spin distribution functions, and $\omega_{\rm s}$ and $\omega_{\rm d}$ 
are the state densities for $\rho_{\rm sph}$ and $\rho_{\rm def}$, respectively. 
The rotational collective enhancement is explicitly treated 
in $\rho_{\rm def}$ by applying the enhancement factor $K_{\rm rot}$.
Contrary to the rotational collective enhancement, 
vibrational one is not explicitly treated in our formulation.
We expect that it is implicitly taken into account through the optimization of the level density parameters.

The level density parameters $a_{\rm s,d}$ are given as, 
\begin{eqnarray} \label{eq_a}
a_{\rm s,d}(U)=a_{\rm s,d}(*)\left[ 1+\frac{E_{\rm sh}}{U}(1-e^{-\gamma U}) \right],
\end{eqnarray}
here $a_{\rm s,d}(*)$ are the asymptotic level density parameters described by the systematics, 
\begin{eqnarray} \label{eq_aasym}
a_{\rm s,d}(*)=\alpha_{\rm s,d} A (1 - \beta_{\rm s,d} A^{-1/3}).
\end{eqnarray}
The parameters $\alpha_{\rm s,d}$, $\beta_{\rm s,d}$, and $\gamma$ are optimized using the experimental $D_0$, 
as explained in the next subsection. 
The shell correction energy $E_{\rm sh}$ is  defied as,
\begin{eqnarray}
E_{\rm sh} = M_{\rm exp}-M_{\rm LDM},  
\end{eqnarray}
here the mass formula of Myers and Swiatecki are \cite{21} used for $M_{\rm LDM}$. 
It is noted that the pairing energy systematics in Eq. \ref{pair} 
is consistent with the one used in the calculation of $M_{\rm LDM}$.

The spin distribution function $R_{\rm s,d}(U,J)$ are
\begin{eqnarray}
R_{\rm s,d}(U,J)=\frac{2J+1}{2\sigma_{\rm s,d}^2}\exp\left[ -\frac{(J+1/2)^2}{2\sigma_{\rm s,d}^2} \right],
\end{eqnarray}
here we employ the shell-corrected spin dispersion function of Mughabghab and Dunford \cite{18},
\begin{eqnarray}
\sigma_{\rm s,d}^2 &=& I_0 \frac{\sqrt{a_{\rm s,d} U}}{a_{\rm s,d}(*)}, \\
I_0 &=& \frac{\frac{2}{5}m_0 R^2 A}{(\hbar c)^2}=0.01389A^{5/3} \hspace{2mm} {\rm MeV^{-1}}.
\end{eqnarray}
The rotational enhancement factor $K_{\rm rot}$ is written as,
\begin{eqnarray}
K_{\rm rot} &=& \sigma_{\perp}^2, \\
\sigma_{\perp}^2 &=& I_0(1+\frac{\beta_2}{3}) \sqrt{\frac{U}{a_{\rm d}}}. 
\end{eqnarray}


In the present model, the composite formula of Gilbert and Cameron \cite{2} is used. 
The low excitation energy region below the matching energy 
$E_{\rm m}$ is described by the constant temperature part $\rho_{\rm CT}(E_x,J)$,
\begin{eqnarray}
\rho_{\rm GC}(E_x,J) &=& R_{\rm h}(U,J) \rho_{\rm CT}(E_x) \hspace{1cm} (E_x<E_{\rm m}), \nonumber \\
\rho_{\rm GC}(E_x,J) &=& \rho_{\rm h}(E_x,J) \hspace{1cm} (E_x \ge E_{\rm m}).
\end{eqnarray} 
Here the spin distribution function $R_{\rm h} (U,J)$ is calculated by, 
\begin{eqnarray} \label{spinf}
R_{\rm h}(U,J) &=& \rho_{\rm h} (U,J)/\rho_{\rm h}^{\rm tot} (U), \hspace{2mm} \rho_{\rm h}^{\rm tot}(U)=\sum_{J} \rho_{\rm h} (U,J),
\end{eqnarray}
where $\rho_{\rm CT}$ is given by, 
\begin{eqnarray}
\rho_{\rm CT}(E_x)=\frac{1}{T}\exp\left( \frac{E_x-E_0}{T} \right) ,
\end{eqnarray}
here $E_{0}$ and $T$ are determined from the usual matching condition \cite{2}.
The pairing corrected matching energy $U_{\rm m}=E_x-2\Delta$ (even-even), $E_x-\Delta$ (odd), $E_x$ (odd-odd) 
are given by the simple systematics, 
\begin{eqnarray} \label{ums}
U_{\rm m}^{\rm sys}=pA^{x} ,
\end{eqnarray}
where the mass dependence of the systematics is introduced to fit the $U_{\rm m}$ 
determined to reproduce the experimental discrete level numbers. 
The optimization procedures for the parameters $p,x$ are explained later.

If the pairing corrected energy $U$ is smaller than 0, the spin distribution function $R_{\rm h}(U,J)$ cannot be calculated by Eq. \ref{spinf}.
To avoid this, we simply extrapolate $R_{\rm h}(U,J)$ at $U=1$ MeV to $U<1$ MeV region.  

Finally, we assume the equal parity distribution function, namely
\begin{eqnarray}
\rho_{\rm GC}(E_x,J,\Pi)=\frac{1}{2}\rho_{\rm GC}(E_x,J).
\end{eqnarray} 

\subsection{microscopic nuclear structure calculation} \label{mic}
In the present model, results of the microscopic nuclear structure calculation is utilized to determine 
the transition from the deformed LD to the spherical LD.
We performed the nuclear structure calculation using FTHFB theory,
and derived the most probable deformation $\beta_2$ as a function of the excitation energy.
The excitation energy is calculated using the energy expectation values of the system with the temperature $T$,  
\begin{eqnarray}
E_x=E(T)-E(T=0).
\end{eqnarray} 
The calculation was executed with HFBTHO code \cite{15}, 
where the energy density functional of SkM* \cite{16} was used. 
We employed the surface-volume mixed type pairing interaction with 
the pairing cutoff energy $\epsilon_{cut}=60$ MeV. The neutron and the proton pairing strengths are determined to 
reproduce the experimental pairing gap derived from the three-point mass difference for $^{120}$Sn and $^{138}$Ba, 
which have the proton and neutron closed shells of $Z$=50 and $N$=82, respectively.

\begin{figure} 
\begin{center}
	\includegraphics[scale=0.35,angle=270]{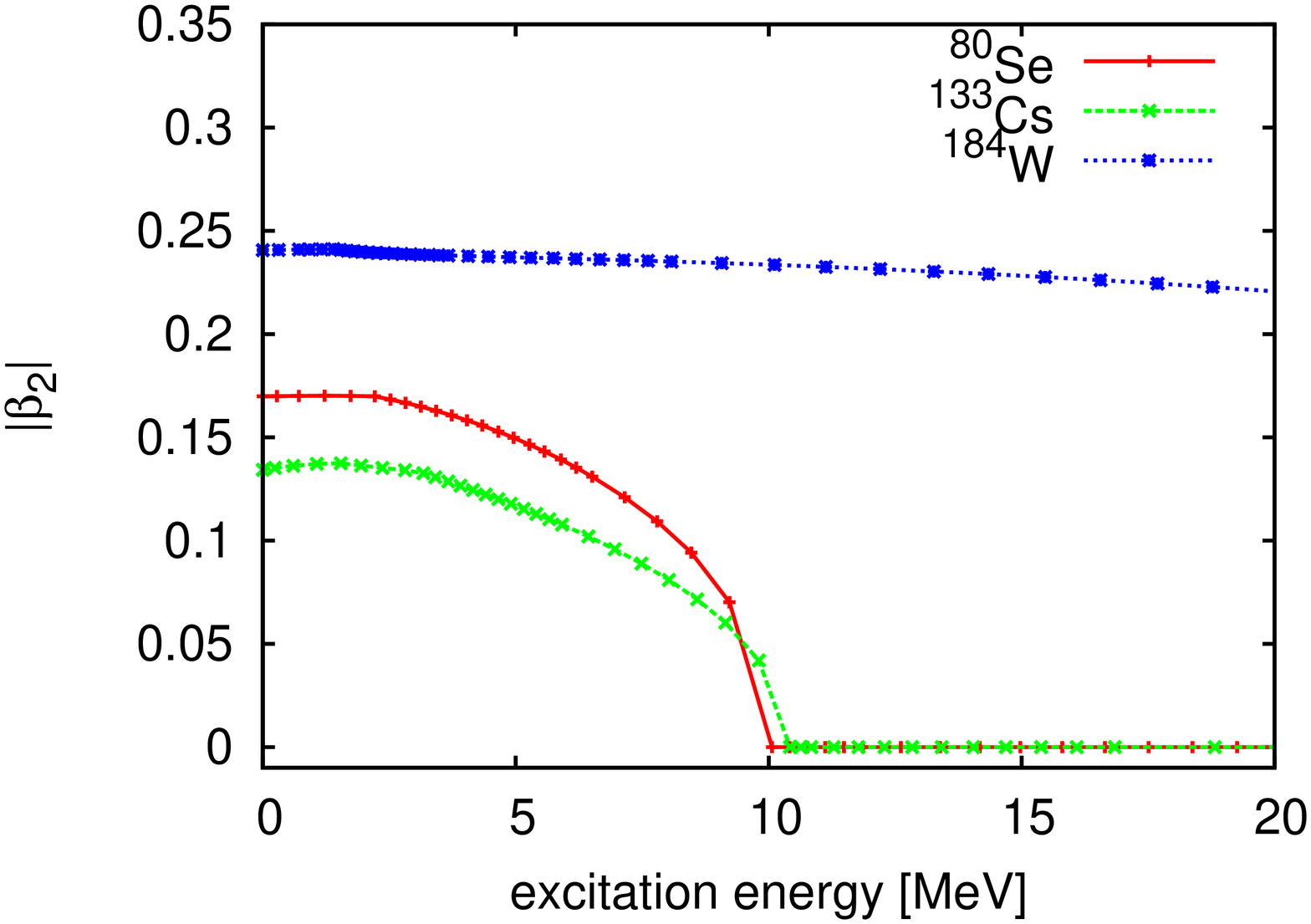} 
	\caption{Most probable deformation $\beta_2$ as a function of the excitation energy calculated by FTHFB.} \label{beta2}
\end{center}
\end{figure}

\begin{figure}
\begin{center}
	\includegraphics[scale=0.35,angle=270]{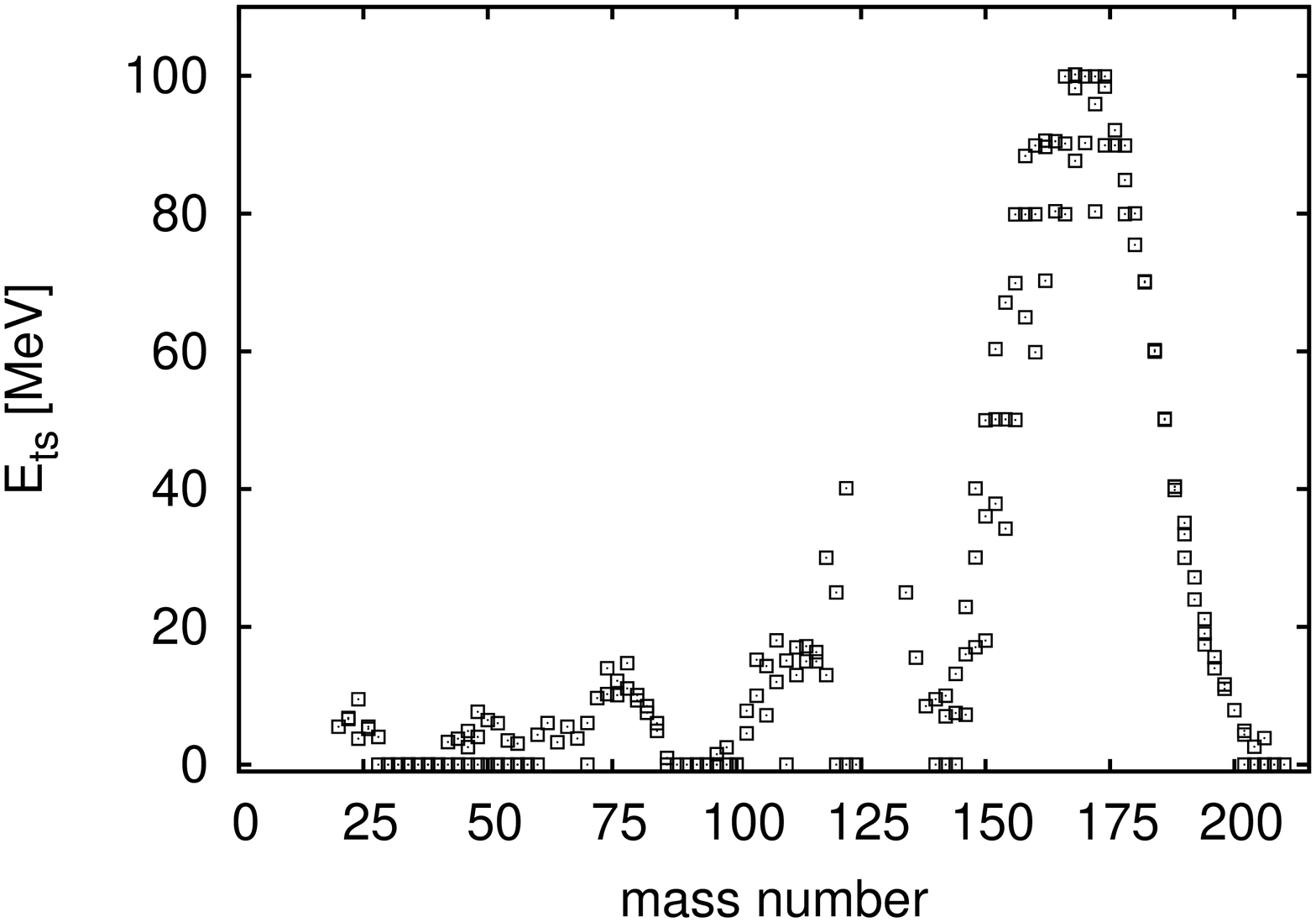} 
	\caption{Parameter $E_{\rm ts}$ derived from FTHFB calculation. } \label{etrans}
\end{center}
\end{figure}

In Fig. \ref{beta2}, the most probable $\beta_2$ as a function of the excitation energy is shown. 
Basically the most probable $\beta_2$ decreases as the excitation energy increases, 
but its behavior is different for each nucleus. For example, 
while $^{80}$Se has a larger $\beta_2$ than $^{133}$Cs at the ground state,
the most probable $\beta_2$ decreases more rapidly and becomes 0 at slightly smaller energy than $^{133}$Cs.  
We define $E_{\rm ts}$ as the energy where the most probable $\beta_2$ value becomes 0, 
because it can be a indicative of the loosing of the rotational collective enhancement, 
and derived it systematically for stable nuclei. The obtained $E_{\rm ts}$ are shown in Fig. \ref{etrans}. 
We found that the most of the deformed nuclei of $A<150$ have $E_{\rm ts}$ of $10\sim20$ MeV. 
This means that the disappearance of the deformation may 
affect nuclear reactions with incident beam energy even below 20 MeV, 
which are often calculated using the statistical model for nuclear data libraries.   
For deformed nuclei in $A>150$ region, the most of them have large $E_{\rm ts}$ which are well above the maximum excitation energy 
of the compound nucleus formed with 20 MeV incident nucleon.   

In the present model, we suppose that the spherical states  
appear in the excited state after exhausting the deformation energy that is defined as the energy difference
between the spherical and the deformed ground state energies,
\begin{eqnarray}
E_{\rm def}=E_{\rm const.}^{\beta_2 =0}(T=0)-E(T=0).
\end{eqnarray}
This energy is subtracted from the excitation energy of $\rho_{\rm sph} (U,J) $, as described by Eq. \ref{eq_cond}.

\subsection{effective and collective level density models}
For comparison, we also derive the LDs using the effective and collective models.
The effective model is defined with $\rho_{\rm sph} (U,J)$ used in the present hybrid model, 
\begin{eqnarray}
\rho_{\rm eff} (U,J)=\rho_{\rm sph}(U,J),
\end{eqnarray}
and the collective model is defined as,
\begin{eqnarray}
\rho_{\rm col} (U,J) &=& \max(\left[K_{\rm rot}-1\right]f(E_x)+1,1)R_d(U,J)\frac{\omega_{\rm d}(U)}{\sqrt{2\pi}\sigma_{\rm d}}, \nonumber \\
f_{\rm dam}(E_x) &=&  \frac{1}{1+e^{(E_x-E_{\rm col})/d_{\rm col}}} , 
\end{eqnarray}
here $E_{\rm col}$ and ${d_{\rm col}}$ are fixed at 30 MeV and 5 MeV, 
which are the values used by Koning et al. \cite{4}.
For both $\rho_{\rm eff} (U,J)$ and $\rho_{\rm col} (U,J)$, the constant temperature 
part are combined in the same way as the hybrid model.

\subsection{optimization procedure} \label{opt}

Basically the optimization of the systematics for the asymptotic level density parameter was performed 
in a similar way to Mengoni and Nakajima \cite{3}. It is noted that the constant temperature model is not used in  
the optimization procedure for the asymptotic level density parameters for simplicity. 

The parameters to be optimized using the experimental values of the average s-wave neutron resonance spacing $D_0$
are $\alpha_{\rm s,d}$, $\beta_{\rm s,d}$ and $\gamma$ in Eq. \ref{eq_a} and \ref{eq_aasym}.
We determine $\alpha_{\rm s,d}$ and $\beta_{\rm s,d}$ to minimize $\chi^2_a$ defined as, 
\begin{eqnarray} \label{chisq}
	\chi^2_a=\Sigma_i \frac{(a_i^{\rm local}(*)-a_i^{\rm sys}(*))^2}{a_i^{\rm sys}(*)},
\end{eqnarray}
here $a_i^{\rm local}(*)$ is the asymptotic level density parameter derived to reproduce the experimental $D_0$ for each nucleus,
and $a_i^{\rm sys}(*)$ is that calculated by Eq. \ref{eq_aasym}. Here $i$ is the index to specify nucleus. 
The experimental $D_0$ values for 300 nuclei are taken from RIPL-3 database \cite{17}.
Once $\alpha_{\rm s,d}$ and $\beta_{\rm s,d}$ are determined,  we calculate $f_{\rm rms}^{D_0}$ defined as,
\begin{eqnarray}
f_{\rm rms}^{D_0}= \exp \left[ \frac{1}{N_{\rm max}} \sum_{i=1}^{N_{\rm max}} {\rm ln}^2 \frac{D_0({\rm cal.})}{D_0({\rm exp}.)}\right]^{1/2},
\end{eqnarray}
where $D_0 ({\rm cal.})$ are calculated using $a^{\rm sys}(*)$.
The above procedure is performed using various $\gamma$ parameters, and finally 
the set of $\alpha_{\rm s,d}$, $\beta_{\rm s,d}$ and $\gamma$ that gives the minimum value of $f_{\rm rms}^{D_0}$ is determined.
Obtained parameters and $f_{\rm rms}^{D_0}$ are listed in Table \ref{tab_as}.

\begin{table} 
	\caption{Parameters of the hybrid, effective and collective models, and calculated  $f_{\rm rms}^{D_0}$.}
	\begin{center}
		\begin{tabular}{cccc} 
			\hline
			& hybrid & effective & collective \\
			\hline
			$f_{\rm rms}^{D_0}$ & 1.66        & 1.74      & 1.66      \\
			$\alpha_{\rm s}$ [MeV$^{-1}$]   & 0.07110 & 0.06573 &           \\
			$\alpha_{\rm d}$ [MeV$^{-1}$]   & 0.01291 &               & 0.03960 \\
			$\beta_{\rm s}$     & -3.608    & -4.385  &           \\
			$\beta_{\rm d}$     & -30.54   &              & -5.708  \\
			$\gamma$ [MeV$^{-1}$]     & 0.072     & 0.073     & 0.098     \\
			$p$ [MeV]            & 547       & 55         &  76    \\
			$x$                     & -1.10        & -0.54     &  -0.74     \\
			$E_{\rm cut}$ [MeV] & 0.30       &             &       \\
			$C$                    & 0.35         &             &       \\
			\hline
		\end{tabular} \label{tab_as}
	\end{center}
\end{table} 

\begin{figure}
\begin{center}
	\includegraphics[scale=0.95,angle=270]{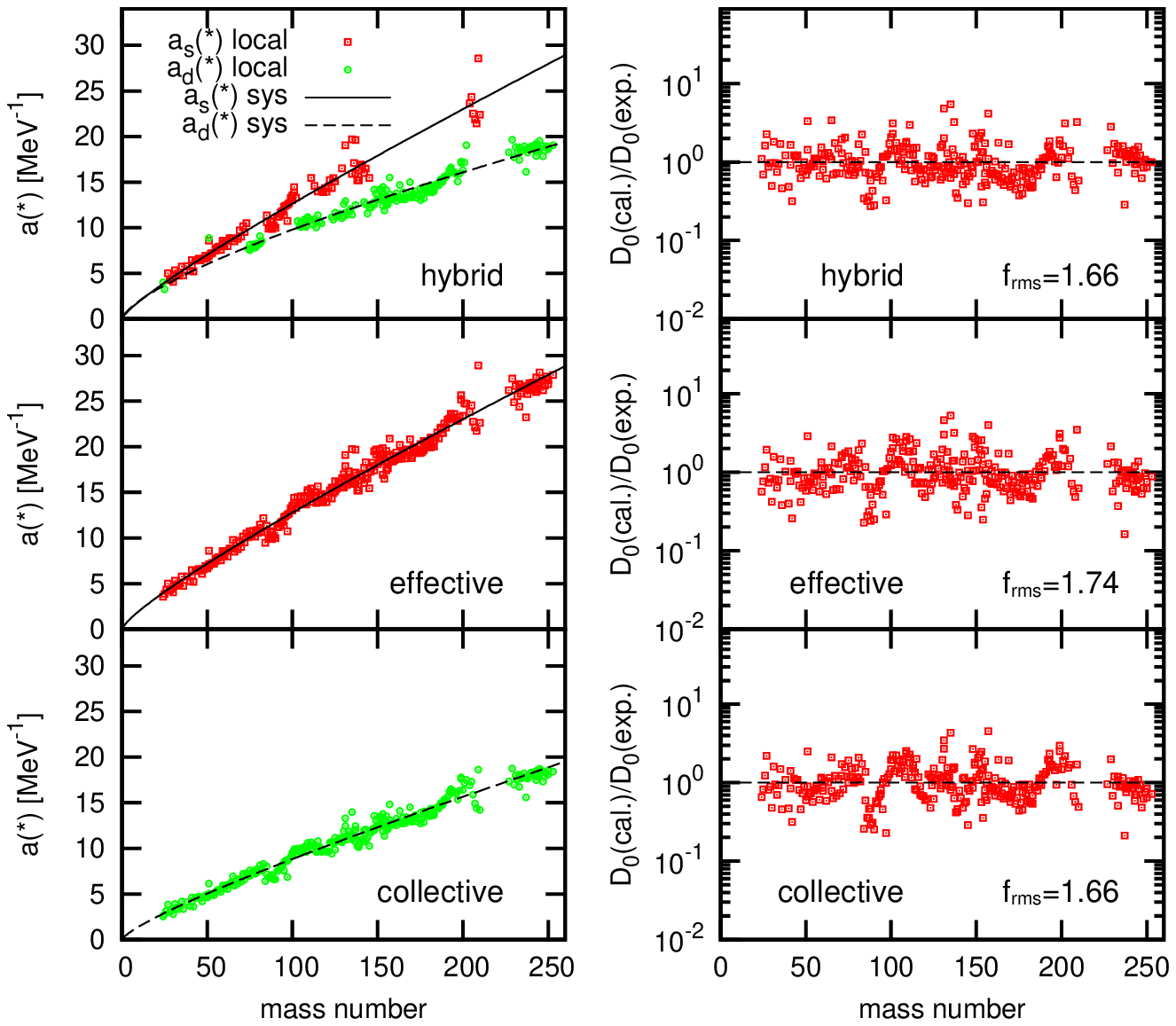} 
	\caption{Calculated a$(^*)$ (left panel) and $D_0$ (right panel) for the hybrid, effective and collective models.  The a$(^*)$ determined to reproduce $D_0$ of each nucleus and calculated from the systematics 
are shown by the symbols and lines, respectively. } \label{d0}
\end{center}
\end{figure}

In more detail, the procedure to determine $\alpha_{\rm s,d}$ and $\beta_{\rm s,d}$ is divided into two steps.
First we determine $\alpha_{\rm s}$ and $\beta_{\rm s}$. 
For the  spherical nuclei that have the condition $E_{\rm def}<E_{\rm cut}$, $D_0$ is calculated only from $\rho_{\rm sph}$. 
Therefore, $a_{\rm s}^{\rm local}$ can be determined independently from $a_{\rm d}$.
In the left top panel of Fig. \ref{d0}, $a_{\rm s}^{\rm local}(*)$ of 108 nuclei with  $E_{\rm def}<E_{\rm cut}$ are shown by the open squares, 
and $a_{\rm s}^{\rm sys}(*)$ determined by minimizing $\chi^2_a$ with these $a_{\rm s}^{\rm local}(*)$ is shown by the solid line.
Secondly, $\alpha_{\rm d}$ and $\beta_{\rm d}$ are determined.
To calculate $D_0$ for nuclei with $E_{\rm def} \ge E_{\rm cut}$, both $a_{\rm s}(*)$ and $a_{\rm d} (*)$ are necessary. 
We calculate $a_{\rm s}(*)$ using $a_{\rm s}^{\rm sys}(*)$ determined from the above procedure, 
and derive $a_{\rm d}^{\rm local}(*)$ to reproduce the experimental $D_0$ 
for 182 nuclei with $E_{\rm def} \ge E_{\rm cut}$.  
The obtained $a_{\rm d}^{\rm local}(*)$ and $a_{\rm d}^{\rm sys}(*)$ 
are shown by the open circles and the broken line in the left top panel of Fig. \ref{d0}, respectively.
It is clearly seen that smaller  $a_{\rm d}(*)$ values are required compared to $a_{\rm s}(*)$, which indicates that 
the spherical and the deformed intrinsic states should have different state densities. 
It is noted that we excluded 10 nuclei with small deformations of 
$E_{\rm cut}<E_{\rm def}<0.5$ MeV, in which $\rho_{\rm h}$ is dominated by $\rho_{\rm sph}$. In such a case, extremely large or small values 
of $a_{\rm d}^{\rm local}(*)$ appears to reproduce $D_0$, and it is unfavorable for the optimization of $a_{\rm d}^{\rm sys}(*)$.

\begin{figure}
\begin{center}
	\includegraphics[scale=0.3,angle=270]{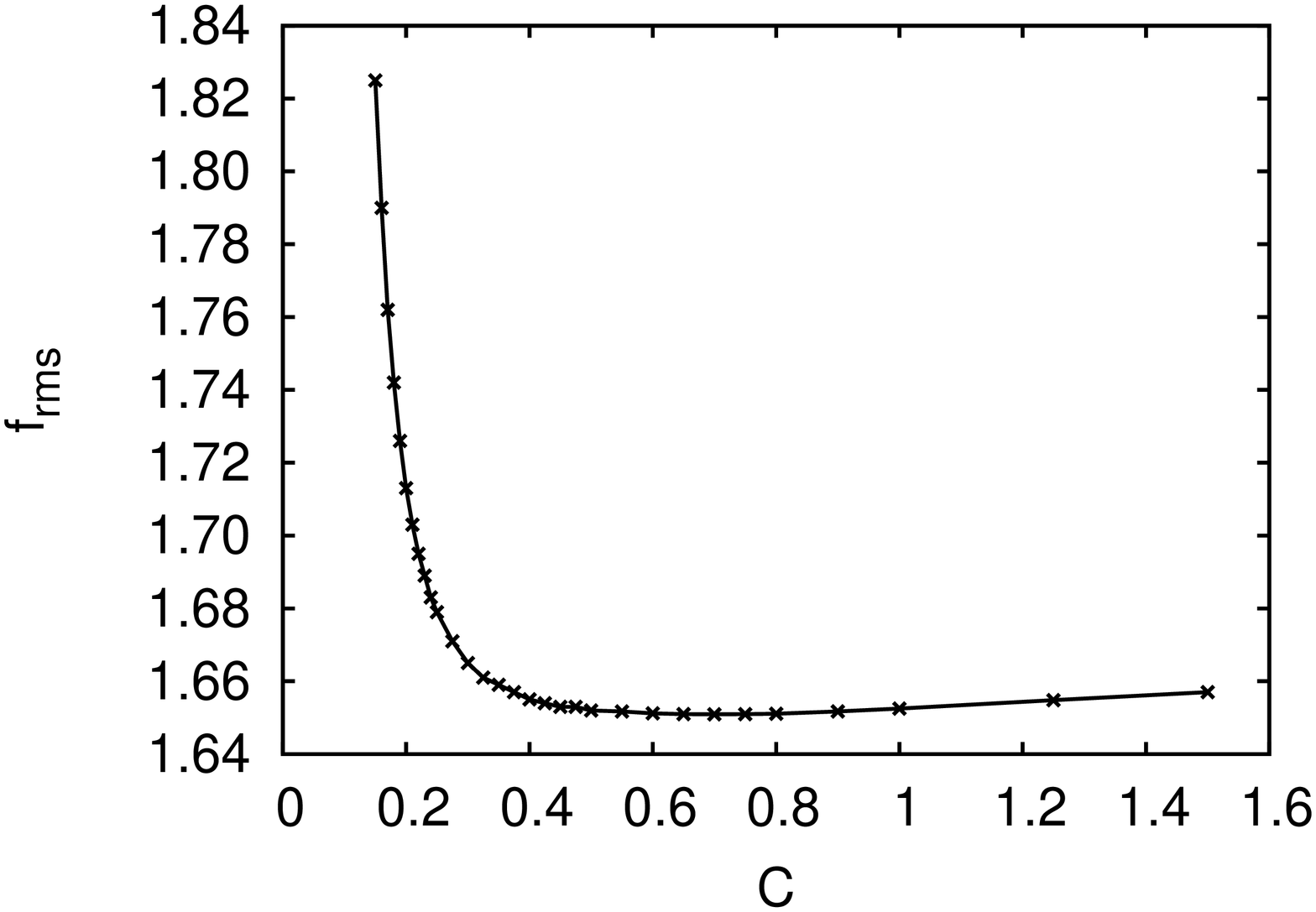} \\
	\caption{Dependence of $f_{\rm rms}^{D_0}$ on the additional parameter $C$ for the hybrid model.} \label{cdep}
\end{center}
\end{figure}

The hybrid model has an additional parameter $C$ that adjusts the width parameter $d_e$ of $f_{\rm dam}$.
While we use the theoretical values for the central energy $E_{\rm ts}$ of $f_{\rm dam}$,  
the width parameter $d_e$ that express a smoothness of the transition is quite phenomenological.
Therefore, we investigated the dependence on $C$ in the calculation of  $D_0$. 
In Fig. \ref{cdep}, $f_{\rm rms}^{D_0}$ as a function of $C$ is shown.
While it is clear that a small $C$ is not preferable, $C$ dependence of $f_{\rm rms}^{D_0}$ is so moderate in lager $C$ region,  
which means that $D_0$ cannot be an strong constraint on $C$.
Basically we used $C=0.35$ that is smaller than the optimal value for $D_0$ that is around 0.70,
since a better agreement between calculations and experimental data of 
the nuclear reaction cross sections was obtained with $C=0.35$, 
in the case of (n,2n) reactions for Se isotopes discussed in the next section.

We also optimized the parameters for the effective and the collective LD models.
For these models, all the experimental $D_0$ values for 300 nuclei are used for the optimization of $a^{\rm sys}(*)$.
The obtained $a^{\rm local}(*)$ and $a^{\rm sys}(*)$ for the effective and the collective models are shown in the middle and the bottom 
panels of Fig. \ref{d0}, respectively, and the parameters in $a^{\rm sys}(*)$ and $f_{\rm rms}^{D_0}$ calculated 
using the optimized $a^{\rm sys}(*)$ are listed in Table \ref{tab_as}. 
Although significantly different parameters are required for $a_{\rm s}^{\rm sys}(*)$ and $a_{\rm d}^{\rm sys}(*)$, 
the resulting $f_{\rm rms}^{D_0}$ are similar among the effective, collective and hybrid models.
As already mentioned in the introduction, the essentiality of the explicit treatment of the collective enhancement is hardly seen in the calculation of $D_0$, if the phenomenological LD models optimized using the experimental $D_0$ are used.

\begin{figure}
	\begin{center}
	\includegraphics[scale=0.3,angle=270]{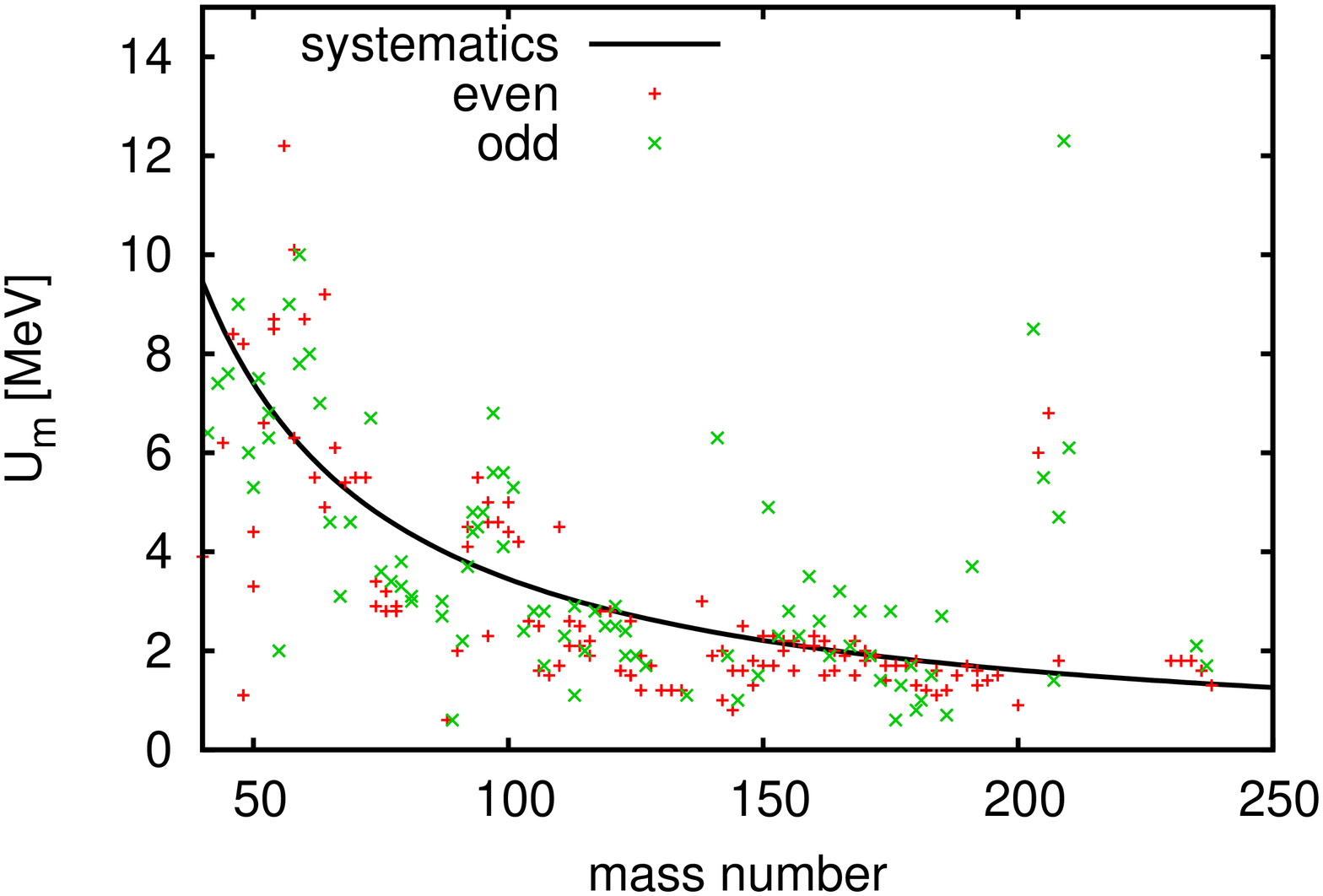} \\
	\caption{Pairing corrected matching energy $U_{\rm m}^{\rm local}$ obtained by minimizing $f_{\rm rms}^{\rm lev}$ of each nucleus and $U_{\rm m}^{\rm sys}$ calculated by Eq. \ref{chisq} 
are shown by the symbols and the solid line, respectively. The red symbols are results for even-even nuclei, and the green ones for odd and odd-odd nuclei.} \label{umatch}
	\end{center}
\end{figure}

Finally, the parameters in the constant temperature part of LD are optimized.
The parameters to be optimized are $p$ and $x$ in Eq. \ref{ums} to calculate $U_{\rm m}^{\rm sys}$.
They are determined to minimize $\chi^2$ calculated as same as Eq. \ref{chisq} using $U_{\rm m}^{\rm sys}$ and $U_{\rm m}^{\rm local}$, and $U_{\rm m}^{\rm local}$ is determined to minimize 
\begin{eqnarray}
f_{\rm rms}^{\rm lev}= \exp \left[ \frac{1}{N_{\rm max}} \sum_{i=1}^{N_{\rm max}} {\rm ln}^2 \frac{L_{E_i} (i) ({\rm cal.})}{L_{Ei}(i) ({\rm exp.})}\right]^{1/2}, 
\end{eqnarray}  
here $L_{E_i} (i)$ is the cumulative number of the discrete levels at the excitation energy $E_i$ 
of the experimentally observed $i$-th level, and $N_{\rm max}$ is the number of levels to be compared. 
The experimental data of the discrete levels are taken from RIPL-3 database \cite{17}. 
Since there may be discrete levels that have not been observed, 
the cumulative number of the observed levels is expected to deviate from the reality with increase of the excitation energy.
We assume that the deviation is small if the cumulative number of the observed levels 
is much smaller than the maximum number of the observed levels, and arbitrary take 70\% of the maximum number as $N_{\rm max}$.
Nuclei with more than 100 observed levels are used to determine the parameters of $U_{\rm m}^{\rm sys}$.
In Fig. \ref{umatch}, the obtained $U_{\rm m}^{\rm local}$ and $U_{\rm m}^{\rm sys}$ are shown by the symbols and the solid line, respectively. It is seen that $U_{\rm m}^{\rm local}$ are roughly reproduced by the mass dependence of $U_{\rm m}^{\rm sys}$, except for the values around $A\sim 200$.
We take priority to achieve better precision for $U_{\rm m}$ in $A<200$ region, which are relevant to the 
nuclear reaction calculations in the next section, and excluded $U_{\rm m}^{\rm local}$ in $A>200$ region from the fitting for this preference. In the final results presented in the next section, the optimized $U_{\rm m}^{\rm sys}$ is used to calculate LD. 


\subsection{Nuclear reaction models}
The nuclear reaction calculations have been executed using CCONE code \cite{14}.
The code composed of the optical model, two-component exciton model, 
distorted-wave Born approximation and Hauser-Feshbach statistical model.
As for the optical model, the global optical potential parameters of Koning and Delaroche \cite{29} was used. 
LDs of the hybrid, effective and collective models are adopted to Hauser-Feshbach statistical model 
in CCONE code by using the tabulated numerical data of RIPL-3 format \cite{17}.
\section{results} \label{res}


\subsection{Total level densities}

\begin{figure}
\begin{center}
	\includegraphics[scale=1.0,angle=270]{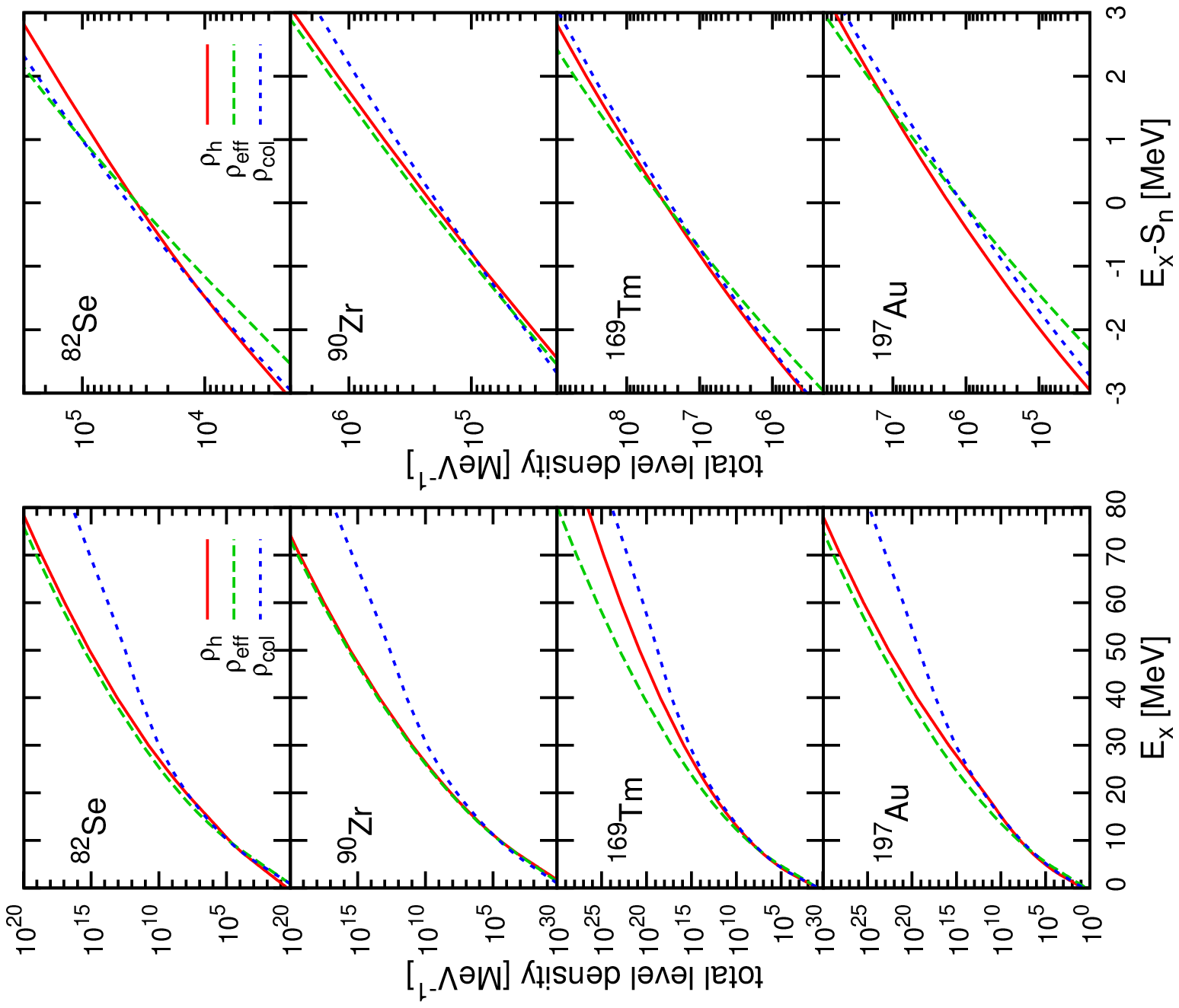}
	\caption{Total level densities of the hybrid (solid line), effective (dashed line) and collective (dotted line) LD models for
$^{82}$Se, $^{90}$Zr, $^{169}$Tm and $^{197}$Au as a function of $E_{\rm x}$ (left panel) and $E_{\rm x}-S_{\rm n}$ (right panel).  } \label{ro}
\end{center}
\end{figure}

Before showing the results of the nuclear reaction calculations, 
the characteristics of the hybrid model are discussed from the total LDs in comparison with the effective and collective models.
In Fig. \ref{ro}, the total LDs of $^{82}$Se, $^{90}$Zr, $^{169}$Tm and $^{197}$Au in wide excitation energy range, and those magnified around the neutron threshold 
are show in the left and right panels, respectively.
The parameters relevant to the deformation that determine the characteristic of the present hybrid model  
are summarized in Table \ref{tab_edef}. 
As described by Eq. \ref{eq_cond} and \ref{eq_fdam}, the transition to $\rho_{\rm sph}$ from $\rho_{\rm def}$ is made by these parameters.
Hereafter, we denote the LDs of the hybrid, effective and collective models 
as $\rho_{\rm h}$, $\rho_{\rm eff}$ and $\rho_{\rm col}$, respectively.

\begin{table} 
	\caption{Calculated $\beta_2$, $E_{\rm def}$, $E_{\rm ts}$ and $E_{\rm m}$ of $^{82}$Se, $^{90}$Zr, $^{169}$Tm and $^{197}$Au.
		 The experimental values of the one neutron separation energies are also shown.}
	\begin{center}
		\begin{tabular}{cccccc} 
			\hline
			           & $\beta_2$ & $E_{\rm def}$ (MeV) & $E_{\rm ts}$ (MeV) & $E_{\rm m}$ & $S_{\rm n}$ (MeV)  \\
			\hline
			$^{82}$Se  & 0.16 & 1.31 & 7.5 & 6.7   & 9.3      \\
			$^{90}$Zr  & 0       &    0   &     0 & 6.2  &  12.0      \\
			$^{169}$Tm & 0.32  & 19.2 & 90.5 & 2.8  & 8.0       \\
			$^{197}$Au  & -0.13  & 3.1 & 13.0 & 2.4  & 6.9       \\
			\hline
		\end{tabular} \label{tab_edef}
	\end{center}
\end{table} 

First of all, for the spherical $^{90}$Zr case, $\rho_{\rm h}$ is close to $\rho_{\rm eff}$ in the entire region, while 
$\rho_{\rm col}$ is significantly different from them, because there is the rotational collective enhancement even 
in the spherical nuclei with the fixed $E_{\rm col}$ of 30 MeV. 
In addition to that, because of the difference in the asymptotic level density parameters, 
the increase rate of $\rho_{\rm col}$ above 30 MeV is also different from $\rho_{\rm h}$ and $\rho_{\rm eff}$.  
As for $^{169}$Tm that has a developed deformation with $\beta_2=0.32$, 
$\rho_{\rm h}$ shows a similar behavior to $\rho_{\rm col}$ below about 30 MeV.  
They deviates from each other above 30 MeV, 
because the rotational collective enhancement fades in $\rho_{\rm col}$ around this energy, 
but does not in $\rho_{\rm h}$.
As for $^{82}$Se that has a moderately developed deformation of $\beta_2=0.16$, 
the component of $\rho_{\rm def}$ in $\rho_{\rm h}$ is decreasing around $E_{\rm x} \sim E_{\rm ts}$=7.5 MeV. 
In $E_{x}>20$ MeV, $\rho_{\rm h}$ comes closer to $\rho_{\rm eff}$, because the component of $\rho_{\rm sph}$ dominates in this region.
The difference between $\rho_{\rm h}$ and $\rho_{\rm eff}$ in the asymptotic region is characterized with the energy shift by $E_{\rm def}$.
$^{197}$Au has a smaller $\beta_2$ of 0.13 but has a larger $E_{\rm def}$ than $^{82}$Se. 
The increment of $\rho_{\rm h}$ significantly reduces around $E_{\rm x} \sim E_{\rm ts}$=13 MeV because the difference between 
the spherical LD shifted by $E_{\rm def}$ and the deformed LD is large. Above 20 MeV, the increase rate of $\rho_{\rm h}$ 
comes closer to $\rho_{\rm eff}$, and deviates from $\rho_{\rm col}$.

The LDs around the neutron threshold $S_{\rm n}$ are shown in the right panel of Fig. \ref{ro} as a function of $E_{\rm x}-S_{\rm n}$.
Since the asymptotic LD parameters are optimized for all of $\rho_{\rm h}$, $\rho_{\rm eff}$ and $\rho_{\rm def}$ 
using the experimental $D_0$, they are close to each other at $S_{\rm n}$. However, there is a difference in the increase rate of these LDs.
In any case, $\rho_{\rm eff}$ has a larger increase rate than $\rho_{\rm col}$. 
Whether the increase rate of $\rho_{\rm h}$ is similar to that of $\rho_{\rm eff}$ or $\rho_{\rm col}$ is determined by the deformation.
It is close to $\rho_{\rm eff}$ for the spherical $^{90}$Zr, and $\rho_{\rm col}$ for the deformed $^{169}$Tm and $^{197}$Au.
As for $^{82}$Se, $\rho_{\rm h}$ has even smaller increase rate than $\rho_{\rm col}$, because the component of $\rho_{\rm def}$ 
disappears just around $S_{\rm n}$ in this case. The increase rates of LDs around $S_{\rm n}$ have remarkable influences on the nuclear reaction calculations explained in the next subsection.  

\subsection{cross sections of (n,xn) and (p,xn) reactions}

In this section, we test the effectiveness of LDs and also discuss the role of the rotational collective enhancement 
from the calculations of (n,xn) and (p,xn) reactions.  
The experimental data of the cross sections to be compared are taken from EXFOR \cite{28} throughout this section.

To illustrate the role of the rotational collective enhancement, 
the (n,2n) and (n,3n) reactions with $^{90}$Zr and $^{169}$Tm targets 
that are spherical and deformed, respectively, are calculated.   
In addition to that, these nuclei have a plenty of (n,2n) experimental data to be compared. 
There are also (n,3n) experimental data for $^{169}$Tm, but not for $^{90}$Zr. Instead, the (n,3n) cross sections of $^{89}$Y are calculated. 

\begin{figure}
\begin{center}
	\includegraphics[scale=0.6,angle=270]{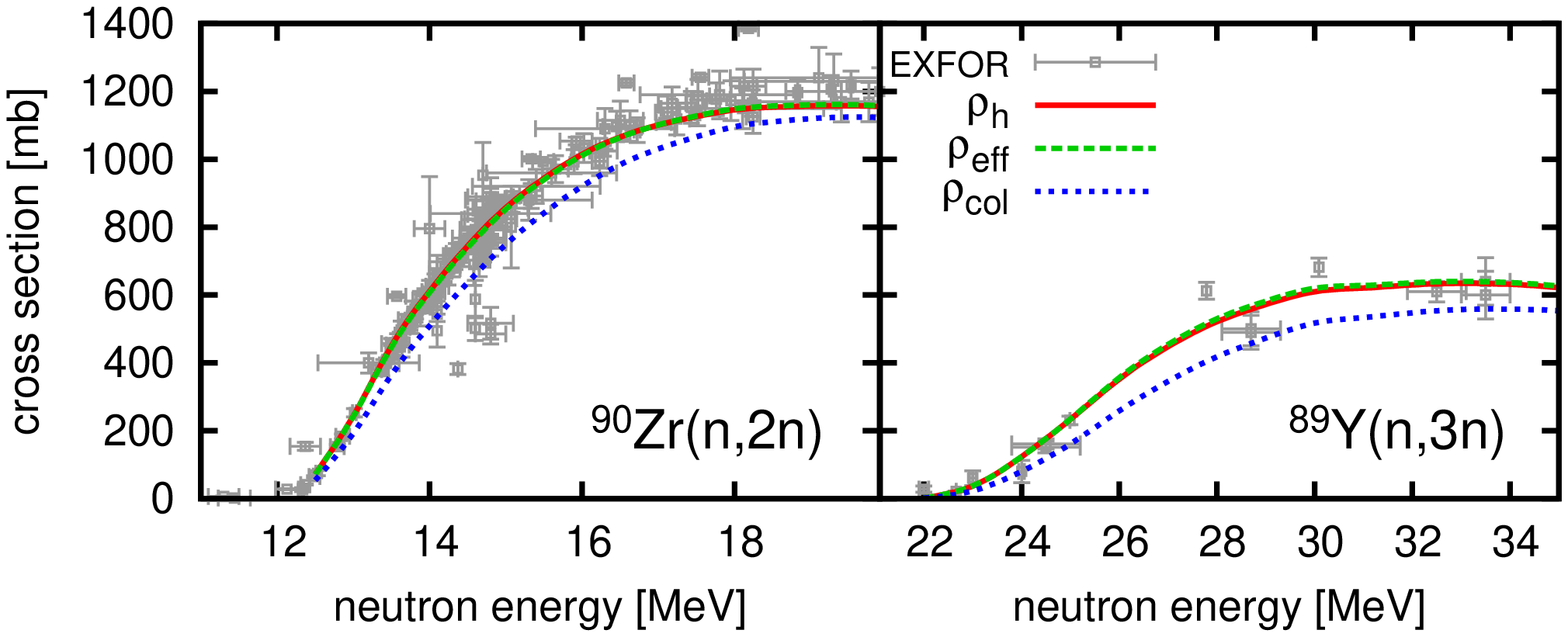} \\
	\includegraphics[scale=0.6,angle=270]{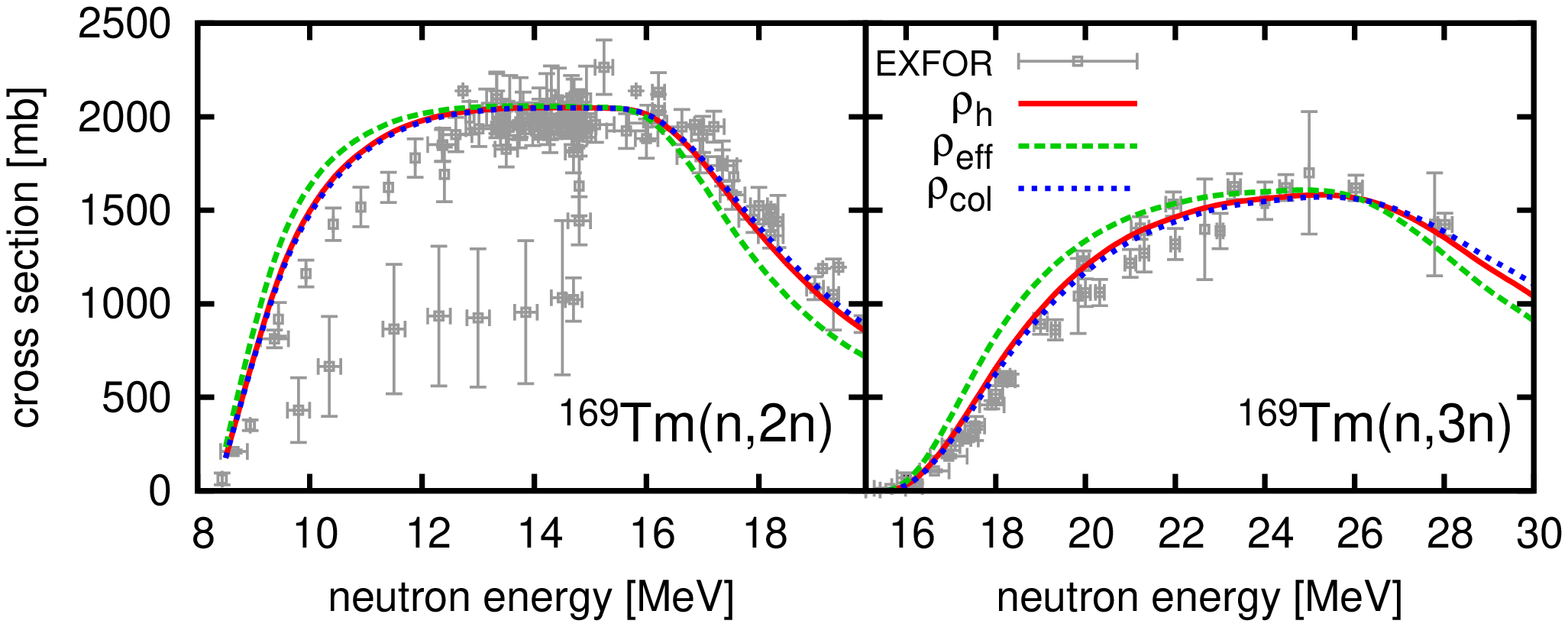}  
	\caption{Cross sections of (n,2n) reactions for $^{90}$Zr and $^{169}$Tm, and (n,3n) reactions for $^{89}$Y and $^{169}$Tm. 
	Calculated results using $\rho_{\rm h}$, $\rho_{\rm eff}$ and $\rho_{\rm col}$ are shown by solid, dashed and dotted lines, respectively.  
They are compared with the experimental data taken from EXFOR shown by symbols.} \label{n2n}
\end{center}
\end{figure}

The results are shown in Fig. \ref{n2n}.
As discussed in the previous subsection, $\rho_{\rm h}$ is similar to $\rho_{\rm eff}$ if the nucleus is spherical. 
Therefore,  for the $^{90}$Zr target, the (n,2n) cross sections calculated using $\rho_{\rm h}$ 
and $\rho_{\rm eff}$ are also similar, and they show good agreement with the experimental data. 
However, $\rho_{\rm col}$ is different from them even for the spherical $^{90}$Zr, and cannot reproduce the experimental data.
On the other hand, for the deformed $^{169}$Tm target, the cross sections calculated with $\rho_{\rm h}$ are similar to those with $\rho_{\rm col}$.
Compared to the results with $\rho_{\rm eff}$, the (n,2n) and (n,3n) cross sections are suppressed below 12 MeV and 25 MeV, respectively.  
The (n,3n) cross sections and the competing (n,2n) cross sections above 15 MeV show good agreement with the experimental data.
The difference in the calculated (n,2n) cross sections mainly come from the difference in the LDs of the target nuclei.
In the (n,2n) reaction, first the $N+1$ compound nucleus is formed, then it emits one neutron. If LD of the target nucleus has smaller
increase rate around the neutron threshold,  the emitted neutron brings more energy, which results 
in the increase of the competitive inelastic channel cross section, 
and decrease of the (n,2n) cross section. 
Later the difference in the neutron emission spectrum is discussed in detail.

\begin{figure}    
\begin{center}
    \includegraphics[scale=0.6,angle=270]{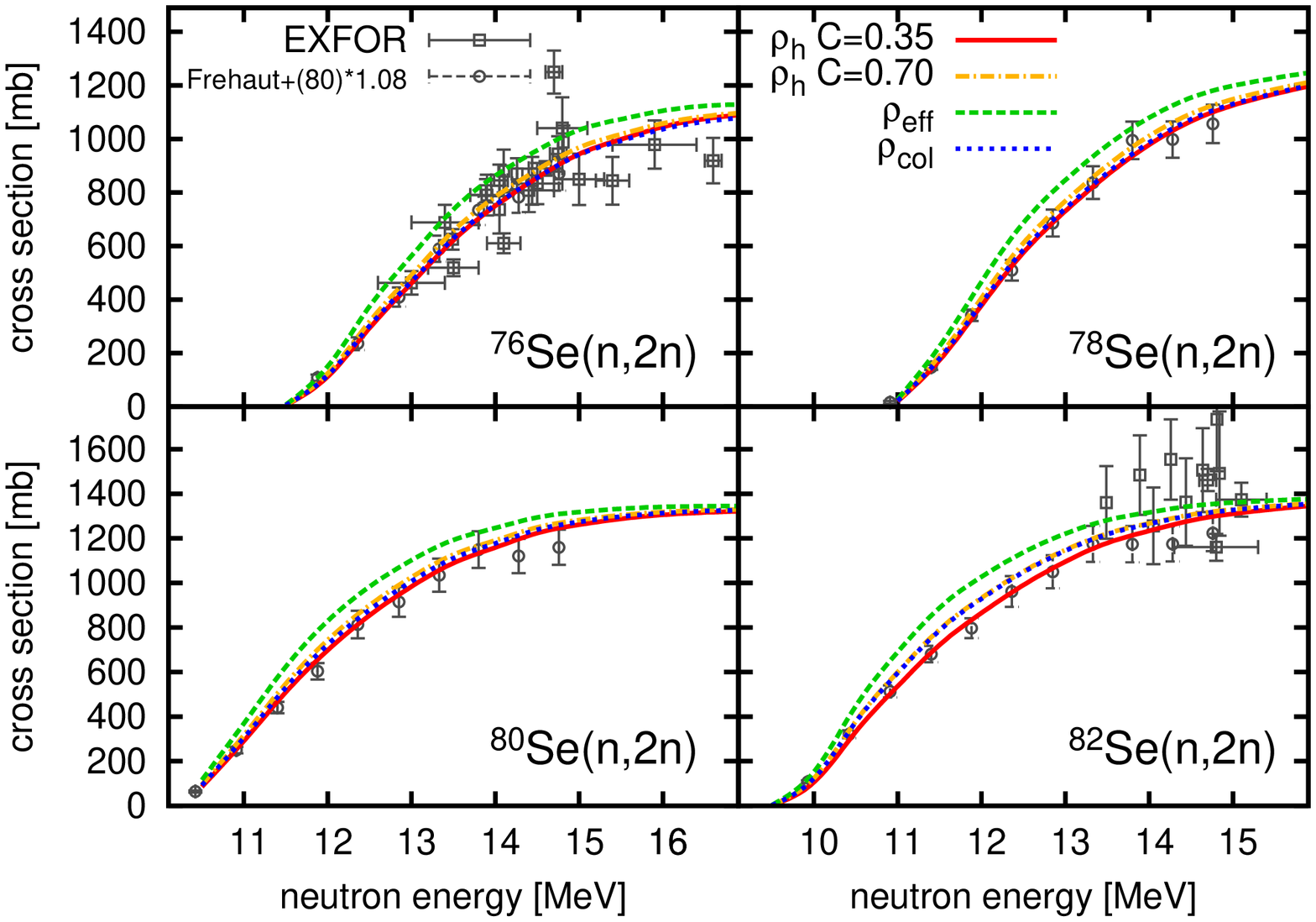} 
	\caption{Cross sections of (n,2n) reactions for Se isotopes. Calculated results are same as in Fig. \ref{n2n} except for the result using $\rho_{\rm h}$ with $C=$0.70 shown by dash-dotted line.
The experimental data of Frehaut et al. are renormalized by a factor of 1.08 \cite{31} (circle).} \label{n2n_se}
\end{center}
\end{figure}

Next we discuss the (n,2n) cross sections of Se isotopes shown in Fig. \ref{n2n_se}. 
If the target nucleus have a moderate deformation with $E_{\rm ts}$ close to $S_{\rm n}$, the (n,2n) cross section calculated with $\rho_{\rm h}$ 
shows non negligible dependence on $d_e$, which is the width parameter of $f_{\rm dam}$.   
$^{76}$Se, $^{78}$Se, $^{80}$Se and $^{82}$Se have $E_{\rm ts}$=12.2, 11.1, 10.1 and 7.5 MeV, and $S_{\rm n}$=11.1, 10.5, 9.9 and 9.3 MeV, respectively.
The (n,2n) cross sections calculated with $\rho_{\rm h}$ and $\rho_{\rm col}$ show suppression from those with $\rho_{\rm eff}$, 
as in the cases of $^{90}$Zr and $^{169}$Tm.
As for the results with $\rho_{\rm h}$, the degrees of the suppression depend on $d_e$.
The results calculated using $C=$ 0.35 and 0.70 are also compared in Fig. \ref{n2n_se}. 
If $d_e$ is smaller, a decrease of the component of $\rho_{\rm def}$ in $\rho_{\rm h}$ is more rapid, which results in a smaller increase rate of LD.
Therefore, the (n,2n) cross sections calculated with $C=0.35$ tend to be suppressed compared to those with $C=0.70$.
While this effect is not significant for $^{76}$Se, $^{78}$Se and $^{80}$Se cases, a noticeable difference is found for $^{82}$Se, 
because $^{82}$Se has $E_{\rm ts}$ just below $S_{\rm n}$. In this case, the component of $\rho_{\rm def}$ becomes 0 
just around $S_{\rm n}$ if $C=0.35$ is used, which results in the significantly small increase rate of LD around $S_{\rm n}$ as shown in Fig. \ref{ro}. 
As for $^{82}$Se, the (n,2n) cross sections calculated with $C=0.35$ are even smaller than those calculated with $\rho_{\rm col}$.

These results indicate that the effect of the fading of the rotational collective enhancement around $S_{\rm n}$ 
can be seen in the (n,2n) cross section. 
The validity of this effect should be studied using as many experimental data as possible, 
but not so many (n,2n) experimental data are available for nuclei that have $E_{\rm ts}$ close to $S_{\rm n}$.
Although the number of experiments is limited, Se isotopes have the systematic experimental data of Frehaut et al. \cite{30}. 
The calculated results with $C=0.35$ well agree with those data renormalized by the factor of 1.08, 
which is derived by Vonach et al. \cite{31}.

\begin{figure}
\begin{center}
	\includegraphics[scale=0.6,angle=270]{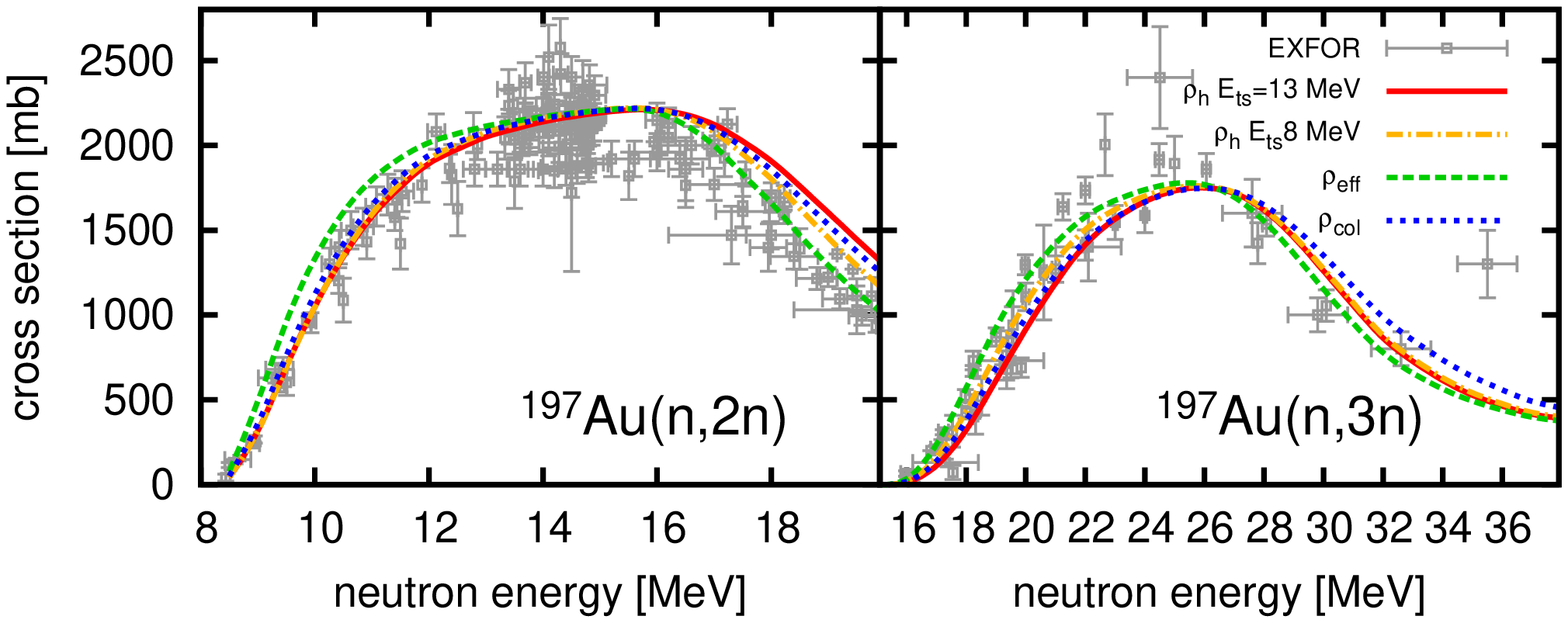} \\
	\includegraphics[scale=0.6,angle=270]{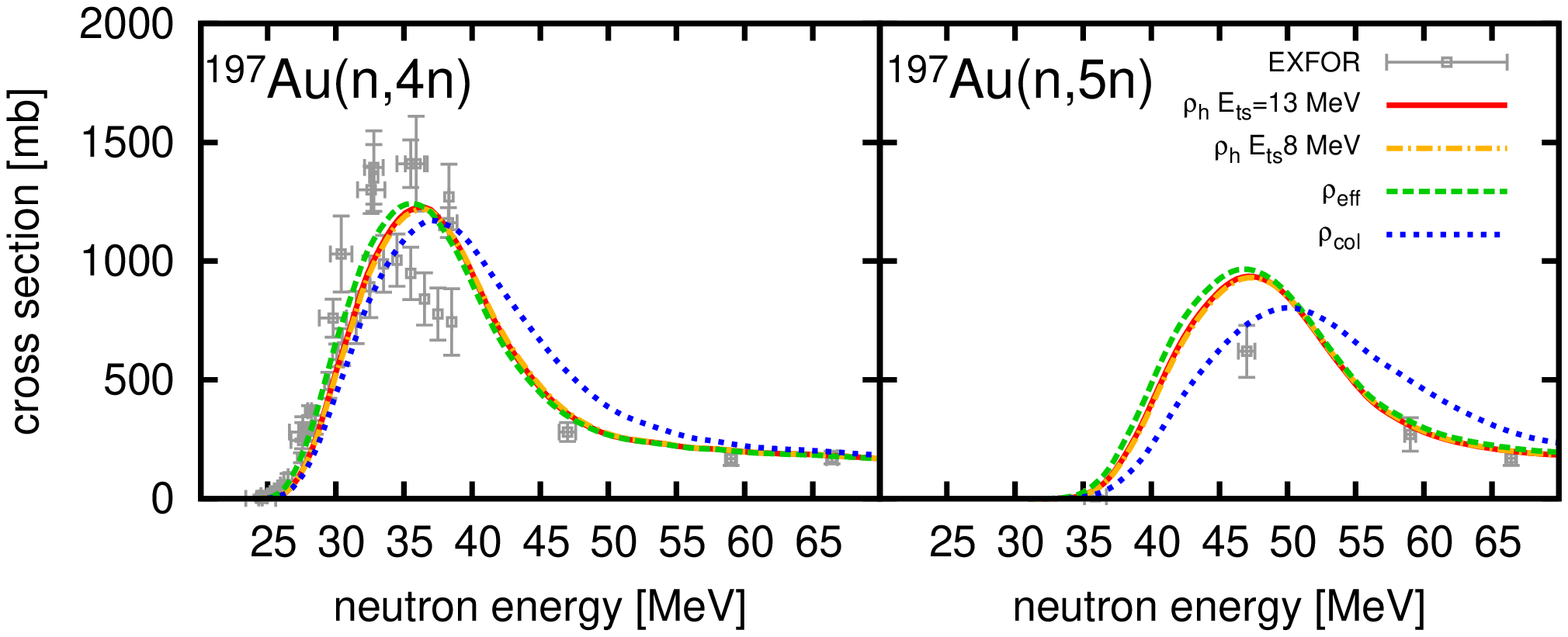}  
	\caption{Cross sections of (n,xn) reactions for $^{197}$Au. Calculated results are same as in Fig. \ref{n2n} except for the result using $\rho_{\rm h}$ with $E_{\rm ts}$= 8 MeV shown by dash-dotted line.} \label{nxn_au}
\end{center}
\end{figure}

Another nucleus that has a plenty of experimental data and a moderate deformation is $^{197}$Au.
The calculated results of $^{197}$Au(n,xn) cross sections are shown in Fig. \ref{nxn_au}.
As discussed in the case of Se isotopes, the values of $E_{\rm ts}$ and $S_{\rm n}$ are important 
to understand the characteristics of the cross section calculated with $\rho_{\rm h}$.
$E_{\rm ts}$ is 13 MeV for $^{197}$Au, while $S_{\rm n}$ and $S_{\rm 2n}$ are 6.6 MeV and 15.0 MeV, respectively.
Since $E_{\rm ts}$ is much larger than $S_{\rm n}$ and just below $S_{\rm 2n}$, both (n,2n) and (n,3n) 
cross sections show suppression from the results with $\rho_{\rm eff}$ below 14 MeV and 25 MeV, respectively. 
However, the (n,2n) and (n,3n) cross sections in 15 MeV $< E_{\rm n} <$ 25 MeV, 
which are competing, show a disagreement with the experimental data.
To investigate how the calculated cross sections depend on the degrees of the deformation, 
a modified $\rho_{\rm h}$ for $^{197}$Au that has arbitrary chosen $E_{\rm ts}$ and $E_{\rm def}$ 
values of 8 MeV and 1 MeV is used to calculate the cross sections.
The results are also shown in Fig. \ref{nxn_au}.
Since $E_{\rm ts}=8$ MeV is well under $S_{\rm 2n}$, the suppression of the (n,3n) cross sections below 25 MeV is small.
As a consequence, this results with the modified $\rho_{\rm h}$ show a better agreement with the experimental data in 15 MeV $< E_{\rm n} <$ 25 MeV.
As for the (n,4n) and (n,5n) cross sections, the results with both $\rho_{\rm h}$ of $E_{\rm ts}$= 8 and 13 MeV are similar, 
because the incident energies are higher enough from $E_{\rm ts}$ for these channels, 
which means the complete disappearance of the component of $\rho_{\rm def}$.
The results with $\rho_{\rm h}$ significantly deviate from those with $\rho_{\rm col}$ in the higher incident 
energy region due to the difference of LDs in the asymptotic region.
Several experimental data above 40 MeV support the results with $\rho_{\rm h}$.
 

\begin{figure}
\begin{center}
    \includegraphics[scale=0.6,angle=270]{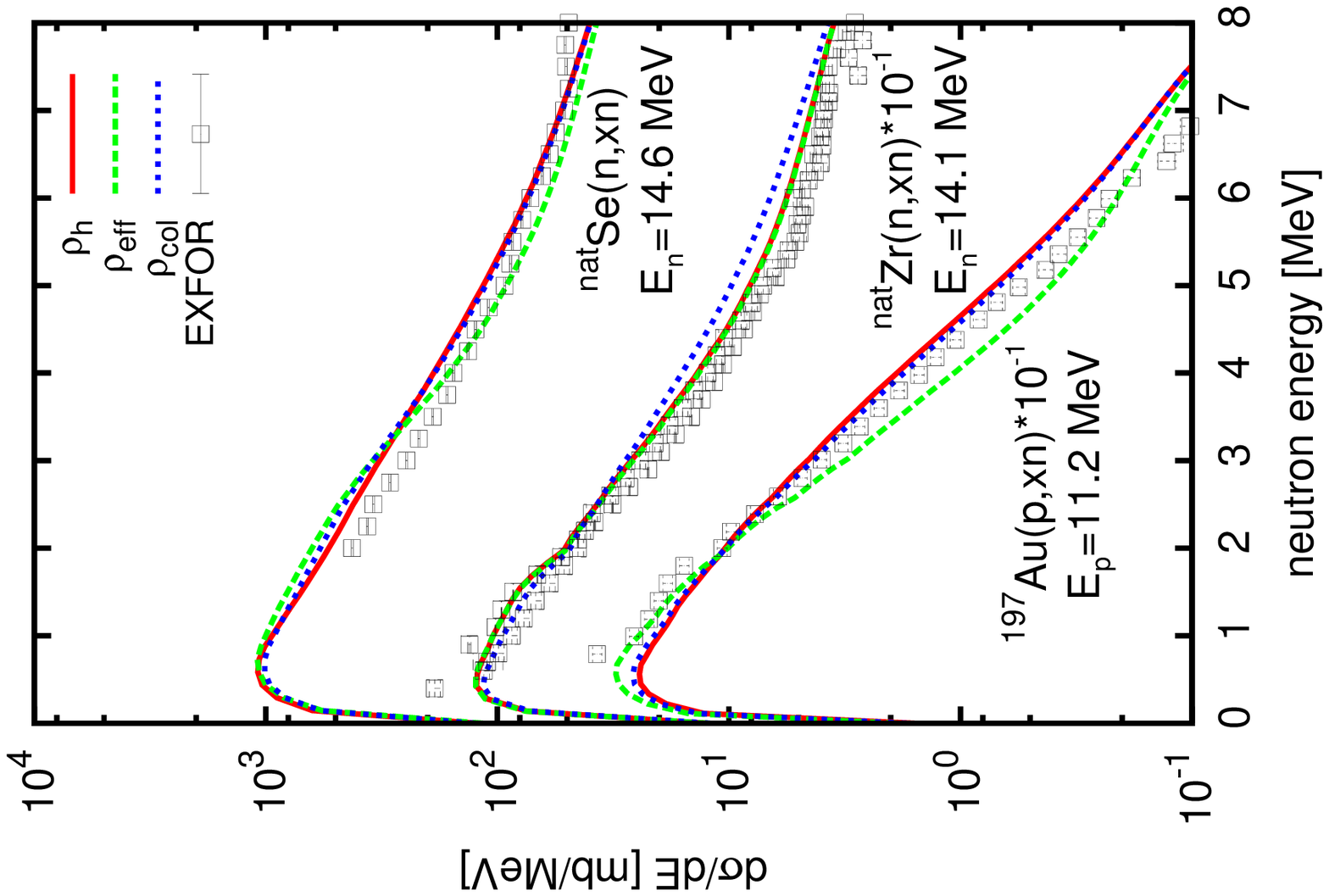} 
    \includegraphics[scale=0.6,angle=270]{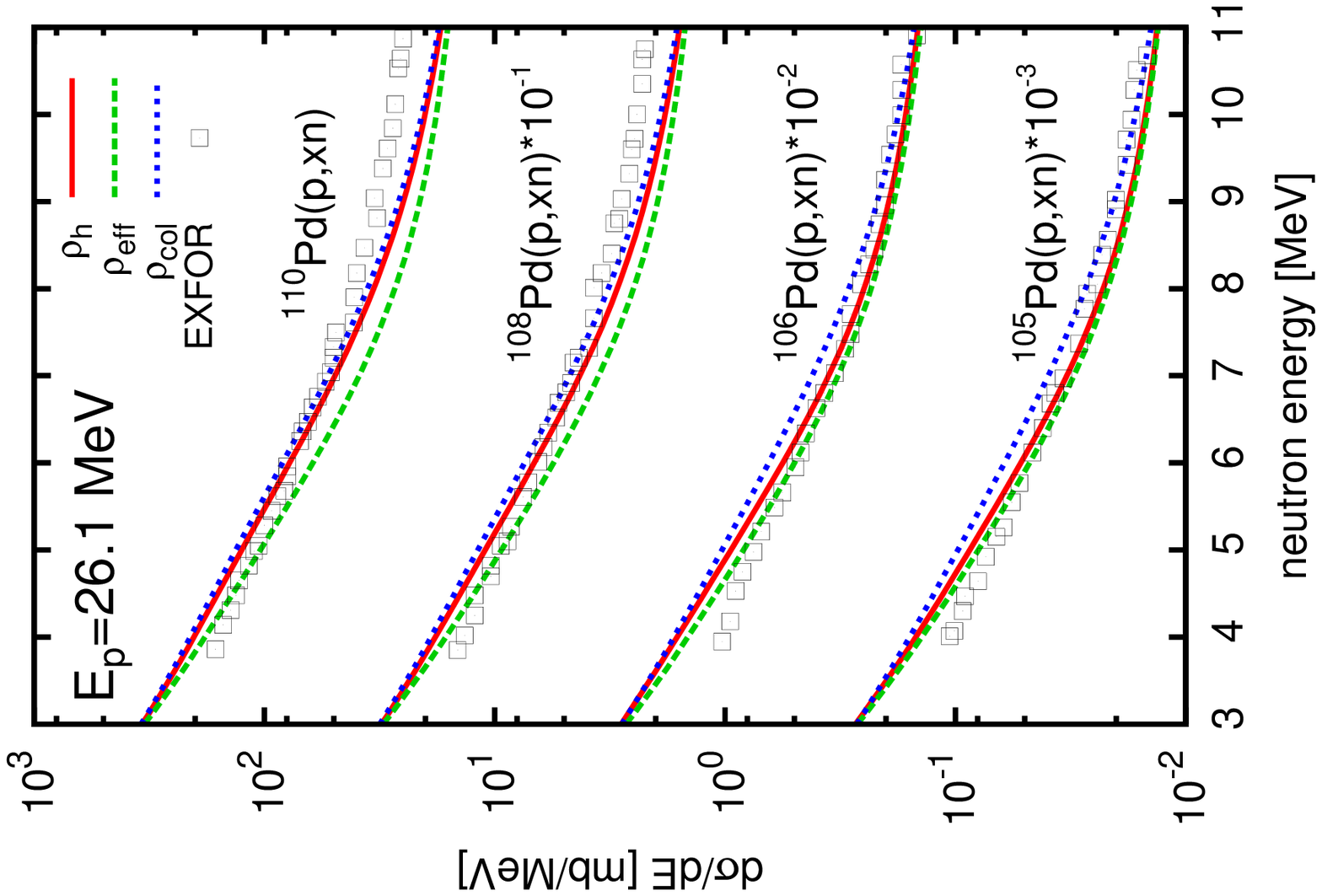} 
    \caption{Neutron emission cross sections of (n,xn) and (p,xn) reactions. Calculated results are same as in Fig. \ref{n2n}. 
The experimental data are taken from EXFOR.} \label{spect}
\end{center}
\end{figure}

The suppression of (n,xn) cross sections calculated 
with $\rho_{\rm h}$ and $\rho_{\rm col}$ from those with $\rho_{\rm eff}$ 
is related to the difference in the evaporated neutron emission spectrum.
To show this, the neutron emission spectrum of $^{\rm nat}$Se(n,xn) $^{\rm nat}$Zr(n,xn) 
and $^{197}$Au(p,xn) reactions are calculated.
The results are shown in the left panel of  Fig. \ref{spect}.
The neutron emission spectrum of $^{\rm nat}$Zr(n,xn) reaction at 14.1 MeV calculated with $\rho_{\rm col}$ 
shows a noticeable enhancement around 5 MeV from those 
calculated with $\rho_{\rm h}$ and $\rho_{\rm eff}$ and a disagreement from the experimental data. 
It is consistent with the (n,2n) cross section calculated with $\rho_{\rm col}$, which significantly deviates from the experimental data. 
Since $\rho_{\rm col}$ has a smaller increase rate at a excitation energy close to the incident nucleon energy,
the evaporated neutrons from the compound nucleus tend to bring larger energies compared to the results with 
$\rho_{\rm h}$ and $\rho_{\rm eff}$. 
In most cases, $\rho_{\rm col}$ has a smaller increase rate than $\rho_{\rm eff}$, even for spherical nuclei.
In $^{\rm nat}$Se(n,xn) case, the calculated result with $\rho_{\rm h}$ is similar to $\rho_{\rm col}$, 
which show enhancement from the result with $\rho_{\rm eff}$ around 5 MeV.  

In the right panel of Fig. \ref{spect}, the neutron emission spectrum of $^{105,106,108,110}$Pd(p,xn) 
reactions at $E_{\rm p}=26.1$ MeV are shown.
For $^{105}$Pd, $^{106}$Pd, $^{108}$Pd and $^{110}$Pd, $E_{\rm ts}$ are calculated to be 
10.0, 11.0, 14.3 18.0 and 20.3 MeV, respectively. 
While all of four Pd isotopes have moderate deformations around $\beta_2 \sim 0.2$, the difference in $E_{\rm ts}$ results in the significant difference in the evaporated neutron emission spectrum.  
Since $^{110}$Pd has the largest $E_{\rm ts}$ that is close to $E_{\rm p}$, the component of $\rho_{\rm def}$ in $\rho_{\rm h}$ affects the neutron emission from the compound nucleus.
In this case, the neutron emission spectrum calculated with $\rho_{\rm h}$ is close to $\rho_{\rm col}$, and deviates from that with $\rho_{\rm eff}$.
If $E_{\rm ts}$ is much smaller than $E_{\rm p}$,  the component of $\rho_{\rm def}$ has a small influence on the neutron emission from the compound nucleus.
Therefore, the neutron emission spectrum calculated with $\rho_{\rm h}$ are similar to those with $\rho_{\rm eff}$ in $^{105}$Pd(p,xn) and $^{106}$Pd(p,xn) cases.
This result illustrates the characteristic of the present LD model, and at the same time, the role of the collective enhancement 
in the evaporated neutron emission spectrum.

 

\section{Summary} \label{sum}
To construct a new phenomenological LD model for a better precision of the nuclear reaction calculation, 
and to investigate the role of the rotational collective enhancement in the nuclear reaction at the same time,
we proposed the hybrid model in which the LDs of the deformed and the spherical states 
described by Fermi-Gas model are connected by the damping function.  
We optimized the asymptotic level density parameter systematics for the LDs of the deformed 
and the spherical states separately using the experimental $D_0$ of deformed and spherical nuclei, respectively.
The information of the nuclear deformation derived from the FTHFB calculation was utilized.
The obtained LD was introduced in the nuclear reaction calculation using the statistical model, 
and the cross sections of (n,xn) and (p,xn) reactions were discussed. 

We found that the LD with the rotational collective enhancement tends to have a smaller increase rate compared 
to that with no explicit collective enhancement, which results in a higher energy neutron emission 
from the compound nucleus. The (n,xn) cross sections with incident neutron energies 
just above the threshold are suppressed because of this mechanism.
In many cases, cross sections calculated with the transitional model were similar to those with the 
effective model and the collective model for the nuclear reactions for the spherical and the deformed targets, respectively.
We showed the calculated examples for the spherical $^{90}$Zr and the deformed $^{169}$Tm targets, both of which agree with the experiments.

Depending on the incident nucleon energy and the degree of the deformation of the target nucleus, 
the cross sections have sensitivity to a certain energy range of LD where the component of the deformed state is decreasing.
In $^{76,78,80,82}$Se(n,2n) reactions, the decreasing component of the deformed state results in 
a good agreement between the calculated and the experimental cross sections. 
In $^{197}$Au(n,xn) reactions, how cross sections depend on the degrees of the deformation was shown.
These results indicate that a more reliable prediction of deformations in excited states may lead to a more precise calculation of cross sections.

These results indicate that the present model is effective for precise calculations 
of nuclear reactions for both the spherical and deformed targets.  
Since the calculated cross section depends on the predicted deformations, 
a more precise cross section calculation can be achieved with a more reliable nuclear structure calculation in future.
This model also can be a tool to investigate the fading of the rotational collective enhancement in nuclear 
excited states through the nuclear reaction calculation. 



\section*{Acknowledgement}
This work was funded by ImPACT Program of Council for Science, 
Technology and Innovation (Cabinet Office, Government of Japan).

%

\end{document}